# Laser Acceleration and Deflection of 96.3 keV Electrons with a Silicon Dielectric Structure

KENNETH J. LEEDLE,[1,*] R. FABIAN PEASE,[1] ROBERT L. BYER,[2] AND JAMES S. HARRIS[1,2]

[1]Department of Electrical Engineering, Stanford University, Stanford, CA 94305, USA

[2]Department of Applied Physics, Stanford University, Stanford, CA 94305, USA

*Corresponding author: kleedle@stanford.edu

**Radio frequency particle accelerators are ubiquitous in ultra-small and ultrafast science, but their size and cost has prompted exploration of compact and scalable alternatives like the dielectric laser accelerator. We present the first demonstration of high gradient laser acceleration and deflection of electrons with a silicon structure. Driven by a five nanojoule, 130 fs mode-locked Ti:Sapphire laser at 907 nm wavelength, our devices achieve accelerating gradients in excess of 200 MeV/m and sub-optical cycle streaking of 96.30 keV electrons. These results pave the way for high gradient silicon dielectric laser accelerators using commercial lasers and sub-femtosecond electron beam experiments.**

Dielectric laser accelerators (DLAs) utilizing commercially available laser systems can support accelerating gradients one to two orders of magnitude higher than conventional radio frequency (RF) accelerators. They also have the capability to operate at the ~10 attosecond (as) timescale [1-5]. Recently, relativistic accelerating gradients of up to 300 MeV/m were demonstrated using $SiO_2$ dual gratings [1], concurrently with subrelativistic 25 MeV/m accelerating gradients using $SiO_2$ single gratings [2,6,7]. These first demonstrations show the potential of dielectric laser accelerators to match or exceed state-of-the-art RF accelerators for a wide range of electron energies. Silicon is an attractive alternative to $SiO_2$ for bridging the accelerator gap between femtosecond photocathode electron sources and attosecond electron applications.

In this letter, we present the first demonstration of high gradient (over 200 MeV/m) acceleration and deflection of electrons with a silicon nanostructure. These results pave the way for attosecond electron beam experiments for ultrafast electron diffraction, time-resolved electron microscopy, and optical streak cameras [8,9]. Applications of relativistic DLA systems include compact radiation therapy devices, x-ray free electron lasers, and TeV energy physics facilities [10].

Silicon provides an excellent platform for future fully integrated accelerator-on-chip systems driven by ultrafast fiber lasers. Precision fiber v-grooves can be etched, low loss and large mode area waveguides can be fabricated on it, and a variety of high coupling efficiency photonic structures can be fabricated for a fully monolithic accelerator system [10]. At relativistic energies, similar laser field coupling efficiencies, $E_z/E_{inc}$, between high index of refraction contrast (e.g. Si/vacuum) and low index contrast structures (e.g. $SiO_2$/vacuum) can be achieved. However, for sub-relativistic applications, the high field coupling efficiencies of silicon structures enable comparable or higher acceleration gradients than with $SiO_2$ structures even though the laser damage threshold is much lower for silicon. The conductivity of silicon also prevents beam steering due to charge accumulation. $SiO_2$ structures require metal coatings or other anti-charging treatment for sub- ~5MeV energies.

A variety of different silicon accelerator structures have been proposed, including woodpile structures [11], photonic crystal slabs [12], and buried grating structures [13]. For our first silicon accelerator demonstration with sub-relativistic 96.3keV electrons, we opted for the simpler reflection-configuration inverse Smith Purcell design. This configuration was first demonstrated with THz radiation and metallic gratings at keV/m accelerating gradients [14]. Breuer and Hommelhoff used a $SiO_2$ inverse Smith Purcell grating in the transmission rather than reflection configuration to achieve their 25 MeV/m accelerating gradient, and required a thin metal coating to prevent charging [2].

Our drive wavelength of 907 nm was a tradeoff between the thermal heating of silicon and the power output of our available Coherent MIRA 900 Ti:Sapphire mode-locked laser. Silicon is an indirect bandgap semiconductor and is only transparent for wavelengths longer than 1.2 µm. However, the absorption length of silicon at 907 nm is 35 µm, so laser fields are minimally attenuated in 200 nm deep gratings [15]. Bulk linear absorption limits our time averaged laser intensity to ~5 MW/cm$^2$ or 70 mJ/cm$^2$ at 76 MHz repetition rate to prevent thermal damage. Ultrafast laser induced breakdown in silicon occurs at approximately 200 mJ/cm$^2$ fluences in the near infrared [16]. Silicon accelerators driven by Er:fiber or Tm:fiber lasers at 1.55 µm and 2 µm in wavelength, respectively, would be subject to two- and three-photon and free-carrier absorption, and could handle significantly higher average powers [17].

The accelerator devices are fabricated from 5-10 ohm-cm phosphorus doped silicon. The grating patterns are written via electron beam lithography using a JEOL JBX-6300 and transferred to the wafer using reactive ion etching. The gratings are 15 µm long and have a period $\Lambda_g$ of 490 nm, which corresponds to the first spatial harmonic $\Lambda_g = \beta\lambda_0$ at an electron velocity $\beta = v/c = 0.54$ and laser wavelength $\lambda_0 = 907$ nm. The grating teeth are 210 nm wide and 200 nm deep, and the transverse magnetic (TM) mode reflectivity is measured to be 8±1%. The gratings sit on a 3 µm high mesa for beam clearance.

We use finite difference time domain (FDTD) simulations to calculate the first harmonic TM mode of the grating. Figure 1 (a) shows the $E_z$ accelerating electric field profile with the net force vectors for all optical phases superimposed. We show the dependence of the acceleration and deflection gradients on the optical phase $\phi$ in Figure 1 (b). The accelerating gradient $G_{acc}$ and deflecting gradient $G_{defl}$ are defined as the average accelerating and deflecting fields for an electron over one grating period:

$$G_{acc} = \frac{1}{\Lambda_g} \int_0^{\Lambda_s} E_z(z(t),t)dz \quad (1)$$

$$G_{defl} = \frac{1}{\Lambda_g} \int_0^{\Lambda_s} \left(E_y(z(t),t) + vB_x(z(t),t)\right)dz. \quad (2)$$

The acceleration and deflection fields are π/2 rad out of phase, and both decay exponentially away from the grating surface with decay constant $\Gamma = \beta\gamma\lambda_0/2\pi = 93$ nm, for $\gamma = (1-\beta^2)^{-1/2}$ [2]. The deflection gradient causes sub-optical cycle streaking of the electrons and the acceleration gradient causes longitudinal energy gain or loss. We measure both accelerating and deflecting forces of a single phase quadrant in our experiment.

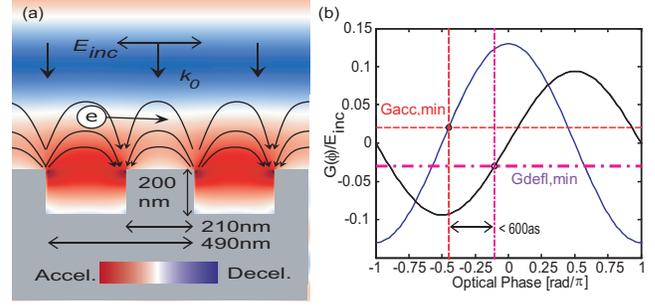

Fig. 1. (a) Accelerating electric field $E_z$ of the grating TM mode at one instance in time from FDTD simulations with force lines for all optical phases superimposed. Electrons experience net acceleration and deflection as they traverse the structure from left to right synchronously with the grating. (b) Acceleration (blue) and deflection (black) gradients $G_{acc}$ and $G_{defl}$ normalized to incident electric field $E_{inc}$ for synchronous electrons vs. optical phase $\phi$. The 600as phase window between our experimental minimum acceleration gradient and minimum deflection gradient is shown.

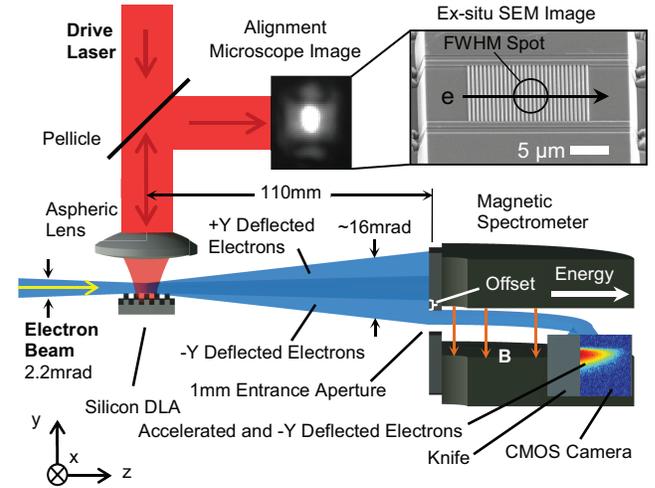

Fig. 2. Experimental setup. 96.3 keV electrons from a scanning electron microscope graze the silicon grating surface, which is illuminated surface normally by a linearly polarized laser. A pellicle beamsplitter picks off ~2% of the back-reflected light to form a microscope image (Ex-situ SEM image of the grating inset). Electrons that are deflected in the –Y direction through the spectrometer entrance aperture are imaged by the CMOS camera if they were accelerated.

Figure 2 shows the experimental setup. A scanning electron microscope (SEM) based on a Kimball Physics EGH-8103 100 keV LaB$_6$ thermal cathode provides a continuous electron beam for the experiment. The electron beam full-angle is 2.2 mrad, and the focal waist radius is 50-100 nm. The nominal electron beam current $I_{tot}$ is 150 pA. An imaging magnetic prism spectrometer [18] mounted on an X-Y stage and CMOS direct detector (Thorlabs DCC1545M with cover glass removed) are used to analyze the electron energy. A knife covers the lowest 30 pixels of the camera and blocks sample-scattered electrons from being detected. The spectrometer entrance has a 1 mm diameter entrance aperture that is used to spatially filter the –Y direction laser deflected electrons from the main beam to further reduce background noise from scattered electrons. Electrons must experience a minimum -Y deflection gradient to enter the spectrometer and a minimum accelerating gradient to be detected by the CMOS camera. This window corresponds to less than 600 attoseconds per 3 femtosecond optical cycle in our experiment. The electron beam angle is determined with a transfer matrix model of the spectrometer and the electron energy by direct comparison against increased cathode potential with no structure present [18]. The spectrometer energy and angular resolutions are approximately 4.5 eV/pixel and 82 μrad/pixel at 96.30 keV, respectively. Measuring both acceleration and deflection simultaneously produces a 100:1 signal to noise ratio.

The Ti:Sapphire laser is focused on the grating using an f = 15.3 mm aspheric lens to a beam waist radius of 3.2±0.3 μm as measured by knife edge. The laser is aligned to the electron beam using the aspheric lens as a microscope objective and an uncoated pellicle beam splitter to create a basic microscope with the sample back-reflected image. Typical laser characteristics for the experiment are: $\lambda_0$ = 907 nm, 400 mW average power, $f_{rep}$ = 76 MHz repetition rate, $\tau_p$ = 130 fs FWHM pulse length, corresponding to a peak fluence of 47 mJ/cm² and peak incident electric field $E_{inc}$ of 1.65 GV/m.

We accumulate electron spectra for 200 seconds for each measurement to increase the signal to noise ratio. The typical accelerated electron count rate for the experiment is $I_{acc}$ = 500-1000 electrons/second, which is in the expected range given the effective electron current $I_{eff}$ = $I_{tot}f_{rep}\tau_p$ ≈ 9200 e/s, the expected detection quantum efficiency of the CMOS camera, and the optical phase window for deflection and acceleration. Space-charge and wakefield effects are expected to be negligible due the low beam current at ~$10^{-4}$ electrons per laser pulse.

For the results shown in Figure 3, the knife edge blocks electrons that gain less than 245 eV, and the spectrometer entrance aperture blocks electrons less than 2.4 mrad off-axis from the electron beam center. We model the accelerated and deflected electron charge density as a 1D Gaussian electron beam centered 150 nm above the grating with a beam waist radius of 65 nm that interacts with the exponentially decaying grating fields. This produces a 'donut' charge density profile in acceleration and deflection phase space with a peak density ring corresponding to the deflection and acceleration experienced at the electron beam center. The model's predicted charge density distribution agrees well with the experimental data as shown in Figure 3 (a).

We integrate over our observed acceleration and deflection phase space to determine the accelerated electron fraction $I_{acc}/I_{eff}$ as a function of energy gain. The model accelerated fraction predictions are compared to the experimental accelerated fraction in Figure 3 (b) and show good agreement. We infer from our simulations that electrons can approach the grating surface within 10 nm. We observe a maximum energy gain of 1.22±0.02 keV before our accelerated fraction drops to the background scattered electron noise level. A maximum beam angle of 7.5 mrad from the beam axis is observed. Using our model and the starting divergence of the beam, we estimate that the peak deflection observed was 5.1±0.5 mrad, for a transverse energy gain of 0.9±0.1 keV.

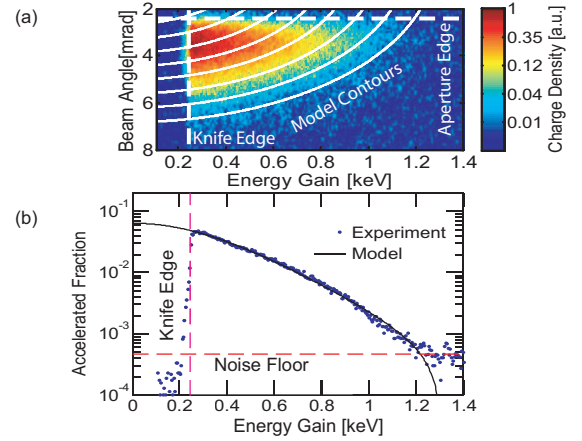

Fig. 3. (a) Example two-hundred second accumulated electron spectrum image for a 2.4 mrad cut-on angle spectrometer offset. Multiply-scattered electrons generate a uniform background noise level over the entire spectrum. Charge density contour lines from the model are superimposed on the spectrum. Energy and angle resolution are 4.5eV and 82 μrad per pixel, respectively. (b) Accelerated fraction $I_{acc}/I_{eff}$ as function of energy gain for the raw spectral data and model fit (line).

Following [6], the characteristic laser-electron interaction distance is

$$L_{int} = \sqrt{\pi}\left(\frac{1}{w_l^2} + \frac{2\ln(2)}{(\beta c \tau_p)^2}\right)^{-1/2} = 5.6 \pm 0.5 \mu m \quad (3)$$

for the laser beam waist radius $w_l$ = 3.2±0.3 μm and $\tau_p$ = 130±5 fs FWHM pulse length. This yields the unloaded acceleration gradient for the 1.22keV observed energy gain

$$G_{acc} = \frac{\Delta U}{L_{int}} = 218 \pm 20 \text{ MeV/m}. \quad (4)$$

The observed acceleration and deflection gradients vary linearly with the incident laser field $E_{inc}$ as shown in Figure 4. This confirms the acceleration field ratio $G_{acc}/E_{inc}$ of 0.13±0.01 and deflection field ratio $G_{defl}/E_{inc}$ of 0.094±0.01, in agreement with FDTD simulations. This is comparable to the acceleration field ratio demonstrated with relativistic $SiO_2$ dual gratings [1]. FDTD simulations indicate that acceleration field ratios of 0.20 can be achieved at $\lambda_0/8$ from a silicon reflection grating as opposed to 0.04 for a comparable $SiO_2$ grating, requiring ~25X less fluence for a given gradient at 96.30keV [2].

Figure 5 (a) shows a $\cos(\theta)$ fit to the acceleration and beam deflection data for the incident laser polarization angle θ and confirms the expected trend.

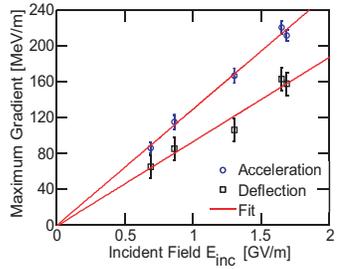

Fig. 4. Maximum acceleration and deflection gradients vs. incident laser field. A linear least squares fit confirms the acceleration field ratio of $G_{acc} = 0.13 \pm 0.01\, E_{inc}$ and deflection field ratio $G_{defl} = 0.094 \pm 0.01\, E_{inc}$.

The accelerating gradient drops off asymmetrically away from the nominal 907 nm drive wavelength, as shown in Figure 5 (b). Accelerated electrons gain momentum from the laser interaction, and they phase-match better with drive wavelengths that are slightly too short. The deflection gradient does not affect the phase matching as much and is nearly symmetric about 907 nm. Particle tracking simulations by numerically integrating the Lorentz force over the full grating FDTD fields agree with the experimental data with no fitting parameters. The experimental energy gains have been normalized for equivalent incident electric field. In future devices, the grating period can be chirped to keep the accelerating field in-phase with the electrons as they gain energy.

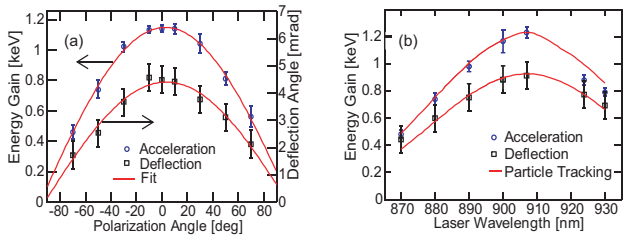

Fig. 5. (a) Accelerator maximum energy gain and maximum deflection angle as a function of drive laser polarization (dots) and $\cos(\theta)$ fit to each curve (lines). (b) Maximum observed longitudinal and transverse energy gain vs. laser wavelength (dots), and the expected curves from particle tracking simulations in our FDTD fields (lines).

Symmetrically illuminated double grating and photonic crystal structures would enable control of the deflection gradients [3,5,11,12,19]. Beam deflection and position control is critical for cascaded structures with longer interaction lengths or for beam steering or undulator applications. Future experiments with femtosecond photocathode electron sources would allow full characterization of the attosecond electron bunches [20].

In summary, we have demonstrated the first silicon-based laser accelerator and deflection structure, with accelerating gradients of over 200 MeV/m with sub-relativistic electrons. We have also measured the deflection characteristics of DLA structures for the first time. The dielectric, electrical, and mechanical properties of silicon, combined with its superb micro-fabrication capabilities, make silicon DLAs well suited to acceleration of electrons for attosecond applications. Single-stage DLAs could be used for optical streaking, compression, or chopping of electron beams at attosecond timescales for ultrafast electron diffraction experiments and time-resolved electron microscopy. Cascaded, chirped DLA structures for compact, high brightness x-ray FELs, radiation therapy devices, and high energy physics could make these technologies accessible at dramatically lower cost and size scales than with RF technology.

**FUNDING INFORMATION**

U. S. Department of Energy (DE-FG02-13ER41970); DARPA (N66001-11-1-4419); AFOSR (FA9550-14-1-0190)

**ACKNOWLEDGMENT**

We thank J. Perales, M. Hennessy, K. Urbanek, E. Wei, and A. Ceballos for technical assistance, and P. Hommelhoff, J. Breuer, R. J. England, O. Solgaard, C. M. Chang, M. Morf, B. Lantz, and D. Waltz for helpful discussions. Devices were fabricated in the Stanford Nanofabrication Facility and Stanford Nano Center.